\documentclass[12pt,a4paper]{article}
\usepackage{epsfig}
\newcommand{\be}{\begin{equation}}
\newcommand{\ee}{\end{equation}}
\newcommand{\bea}{\begin{eqnarray}}

\newcommand{\eea}{\end{eqnarray}}
\newcommand{\nn}{\nonumber}
\newcommand{\dd}{\displaystyle}
\newcommand{\spur}[1]{\not\! #1 \,}
\def\slash#1{\setbox0=\hbox{$#1$}#1\hskip-\wd0\dimen0=5pt\advance
\dimen0 by-\ht0\advance\dimen0 by\dp0\lower0.5\dimen0\hbox
to\wd0{\hss\sl/\/\hss}}
\def\Black{}

\def\Brown{}
\begin{document}
\begin{titlepage}
\title{\hfill 
\begin{center}
\Brown Semileptonic and nonleptonic $B$ decays to three charm quarks:
$B \to J/\psi~(\eta_c)~D~\ell \nu_\ell~{\rm and}~J/\psi~(\eta_c)~D~\pi$ 
\end{center}}

\vspace{+1truecm}

\author{\Black Gad Eilam, Massimo Ladisa and Ya-Dong Yang}

\date{~}
\maketitle

\vspace{-2truecm}
\begin{it}
\begin{center}
Physics Department, Technion-Israel Institute of Technology,
 Haifa 32000,  Israel
\vspace*{0.3cm}
\end{center}
\end{it}
\begin{abstract}
\noindent We evaluate the form factors describing  the semileptonic decays 
$\overline{B^0}\to J/\psi~(\eta_c)~D^+~\ell^- \bar \nu_\ell$,
within the framework of a QCD relativistic potential model. 
This decay is complementary to
$\overline{B^0}\to J/\psi~(\eta_c)~D^+~\pi^-$ 
in a phase space region where a pion factors out.
We estimate the branching ratio for these semileptonic and nonleptonic 
channels, finding 
$\mathcal{BR}(\overline{B^0} \to J/\psi~(\eta_c)~D^+~\ell~\nu_\ell) 
\simeq 10^{-13}$, 
$\mathcal{BR}(\overline{B^0} \to J/\psi~D^+~\pi^-) = 3.1 \times 10^{-8}$ and 
$\mathcal{BR}(\overline{B^0} \to \eta_c~D^+~\pi^-) = 3.5 \times 10^{-8}$.
\end{abstract}\thispagestyle{empty}
\vspace*{3cm}
\end{titlepage}
\setcounter{page}{1}

\noindent In this article we study the $\overline{B^0}$ meson decays 
\bea 
&& 
\overline{B^0} \to J/\psi~(\eta_c)~D^+~\ell^- \bar \nu_\ell  \label{semilep}
\\
&&
\overline{B^0} \to J/\psi~(\eta_c)~D^+\pi^-~. \label{nonleptonic}
\eea

\noindent These decays which may be classified as ``very rare'',
at least in the Standard Model with mesons built solely 
of quark-antiquark pairs, should be searched for in
present and future programs at $B$-factories. As a possible 
motivation consider preliminary 
studies of the inclusive $B\to J/\psi X$ spectrum, 
indicating a slow $J/\psi$ hump \cite{CLEO, Belle}, which kinematically 
corresponds to an invariant mass $m_X \simeq 2$ GeV. Some hypotheses have 
been already suggested in order to account for such a phenomenon: in 
\cite{brod} the $\overline{B^0}\to J/\psi~\Lambda~\bar p$ decay is followed 
by the resonant $\Lambda-\bar p$ bound state, whereas in \cite{hou} a 
possible explanation is the intrinsic charm content of the $B$-meson. In the 
latter case the decay proceeds through the 
$\overline{B^0}(b\bar d (\bar c c)_{\rm slow}) \to 
J/\psi_{\rm slow} D^{\star 0}$ 
channel and it accounts for a $\mathcal{BR}\simeq 10^{-4}$ provided that the 
intrinsic charm content fraction in the incoming $B$-meson, is at least 1\%.
	
To corroborate the hypotheses in \cite{brod, hou}, it is worth estimating the 
mechanisms for these decays in the framework of conventional heavy mesons 
picture as precisely as we can. In \cite{ELY} the 
$\overline{B^0}\to J/\psi~(\eta_c)~D^{(\star) 0}$ decays have been calculated 
in perturbative QCD; the branching ratios for these decays are estimated 
around 10$^{-7}$-10$^{-8}$ and, therefore, too small to account for the 
Belle and CLEO data. Moreover in \cite{ELY} the possibility of production of a hybrid 
$s \bar d g$ meson with mass around 2 GeV is briefly discussed and, although the 
calculation of such a decay is difficult, a decay rate $10^3 \simeq 10^4$ larger than 
the conventional mechanism  for $\mathcal{BR}$ is expected. 

The nonresonant decay mode 
$\overline{B^0} \to J/\psi~(\eta_c)~D^+~\pi^-$
would be interesting to analyze in this context, as it might
provide a significant background to the decay process 
$\overline{B^0} \to J/\psi~(\eta_c)~D^{\star 0}$, followed by
$D^{\star 0} \to D^+~\pi^-$ with a slightly off-shell $D^{\star 0}$.
While a calculation from first principles is not available at the
moment, a useful approximation might be the factorization
approximation \cite{facto} and, within this framework, the decay modes
(\ref{semilep}) 
would provide the crucial
hadronic matrix elements needed to compute the relevant
amplitudes. 

From a theoretical standpoint, semileptonic $B$-meson decays with
two hadrons in the final state represent a formidable challenge as
they involve hadronic matrix elements of weak currents with three
hadrons. They can be studied by pole diagrams, which amounts to a
simplification because only two hadrons are involved in the
hadronic matrix elements. This is the approach followed in some
papers where these decays have been examined  in the framework of
the chiral perturbation theories for heavy meson decays
\cite{lee, burdman}. This method has been successfully applied 
to decays with light mesons in the final state and it is based on an effective
theory implementing both heavy-quark and chiral symmetry
\cite{burdman, wise, Wolfenstein, casalbuoni}. This method
allows to achieve, for systems comprising of both heavy ($Q$) and
light ($q$) quarks, rigorous results  in the combined
$m_Q\to\infty,~m_q\to 0$ limit. However the range of validity of
this approach is limited by the requirement of soft pion momenta and it 
has been never applied, to our knowledge, to a final state with three charm 
quarks. The aim of this article is to
examine the decays (\ref{semilep}) in the
framework of a QCD relativistic potential model \cite{pietroni}
and to extend the kinematical range where theoretical predictions
are possible. We shall present a detailed analysis of the form factors 
relevant for (\ref{semilep}). Subsequently, the decays 
(\ref {nonleptonic}) will be considered.
We do not include  final state interactions
in our calculation since no consistent way to compute them is
presently available. It is clear that they can modify our 
numerical results \cite{nard-pham}.

\par
In three recent papers \cite{Brho, Bpi, Bpipi} an
analysis of some semileptonic and rare $B$-meson decays into one and two light
hadrons has been presented; it employs the relativistic potential model in an
approximation that renders the calculations simpler. We wish to
exploit here this approximation in the study of the 
$B \to J/\psi~(\eta_c)~D~\ell  \nu$ decay.

Let us start with a description of the model (for more details see
\cite{pietroni, Brho, Bpi, Bpipi}). 
In this approach
the mesons are described as bound states of constituent quarks and
antiquarks tied by an instantaneous potential $V(r)$, which  has a
confining linear behaviour at large interquark distances $r$ and a
Coulombic behaviour $\simeq -\alpha_s (r)/r$ at small distances,
with $\alpha_s (r)$ the running strong coupling constant (the
Richardson's potential \cite{Rich}  is used to interpolate between
the two regions).
Due to the nature
of the interquark forces, the light quarks are
relativistic; for this reason one employs for the meson wave
function $\Psi$ the Salpeter \cite{salpeter} equation embodying
the relativistic kinematics:
\be
\left [ \sqrt{ -\nabla_1^2 + m^2_1} + \sqrt{ -\nabla_2^2 + m^2_2}
+V(r)\right ] \Psi(\vec r)=M \Psi(\vec r)~ \label{salp} \;\; , \ee
where the index 1 refers to the heavy quark and the index 2 to the
light antiquark; $M$ is the heavy meson mass that is obtained by
fitting the various parameters of the model, in particular the
heavy quark masses which are fitted to the values $m_c=1452$ MeV,
~$m_b=4890$ MeV, 
and the
light quark masses $m_u\simeq m_d=38$ MeV, $m_s=115$ MeV.
 The heavy meson wave function $\Psi_H(\vec r)$ in its rest
frame is obtained by solving eq. (\ref{salp}) ($H=$ heavy bound state); 
a useful
representation in Fourier momentum space was obtained in \cite{Brho}
and is as follows
\be
\psi (k)=4\pi\sqrt{M \alpha^3}\ e^{-\alpha k} \;\; , \label{wf}
\ee with $\alpha=2.4~(1.6)$ GeV$^{-1}$ for $B,~D~(J/\psi,~\eta_c)$ mesons and 
$k=|\vec k|$ the quark
momentum in the heavy meson rest frame; this is the first approximation
introduced in \cite{Brho}. At this point we would like to comment
that the spectrum obtained depends only weakly on the light quarks 
masses. We will exploit this fact later when we employ 
$m_d=\Lambda_{\rm QCD}$.

The constituent quark picture used in the model is rather crude.
There are no propagating gluons in the instantaneous
approximation {\it i.e.}, the Coulombic interaction is assumed to be static.
Moreover, the complex structure of the hadronic vacuum is
simplified: confinement can be introduced by the linearly
rising potential at large distances, but the chiral symmetry and
the Nambu-Goldstone boson nature of the $\pi$'s cannot be
implemented by the constituent quark picture. For these reasons,
while there are good reasons to believe that eq. (\ref{salp}) may
describe the quark distribution inside the heavy meson, one cannot
pretend to apply it to light mesons. Therefore pion couplings to
the quark degrees of freedom are described by effective vertices 
(for more details see \cite{Brho, Bpi, Bpipi}).

To evaluate the amplitude for semileptonic decays, it is useful to
follow some simple rules, similar to the Feynman rules by which
the amplitudes are computed in perturbative field theory. The
setting of these rules is the main innovation introduced in
\cite{Brho} as compared to \cite{pietroni}. For the decays
(\ref{semilep}) we draw a quark-meson
diagram as in fig. \ref{diagram} and  evaluate it according to
the following rules:
\par\noindent
1) For the heavy meson  $H$ in the initial state one introduces
the matrix:
 \be
\mathcal{H}=\frac{1}{\sqrt3}\psi_H (k){\sqrt{ \frac{m_q m_Q} {m_q m_Q+q_1\cdot
 q_2} }}\;\;\frac{\spur{q_1}+m_Q}{2 m_Q} \Gamma
 \frac{-\spur{q_2}+m_q}{2 m_q} \label{B}
 \ee
where $m_Q$ and $m_q$ are the heavy and light quark masses,
$q^\mu_1,\ q^\mu_2$ their $4-$momenta and 
$\Gamma=~-i \gamma_5,~(\spur{\epsilon})$ for a $J^P=~0^-~(1^-)$ heavy meson. 
The normalization factor
corresponds to the normalization $<H|H>=2\ m_H$ and $\dd\int
\frac{d^3 k}{(2\pi)^3} |\psi_H(k)|^2=2 m_H$ already embodied in
(\ref{B}). One assumes that the $4-$momentum is conserved at the
vertex $H\bar qQ$, i.e. $q^\mu_1+q^\mu_2=p^\mu=$ $H$-meson
$4-$momentum. Therefore $q^\mu_1=(E_Q,\vec k),~q^\mu_2=(E_q,-\vec
k)$ and
\be
E_Q+E_q=m_H \;\; . \label{Alt-Cab} \ee
\noindent
2) For the heavy meson $H$ in the final state one introduces the matrix:
\be
-\gamma^0 \mathcal{H}^\dagger \gamma^0 \; ,
\label{Bbar}
\ee
where $\mathcal{H}$ is as defined in eq. (\ref{B}).
\par
\noindent
3) To take into account the
off-shell effects due to the quarks interacting in the meson, one
introduces running quark mass $m_Q(k)$, to enforce the condition
\be
E=\sqrt{m^2(k)+|\vec k|^2} \label{Alt-Cab2} \ee for the
constituent quarks\footnote{By this choice, the average
$<m_Q(k)>$ does not differ significantly from the value $m_Q$
fitted from the spectrum, see \cite{Brho} for details.}.
\par
\noindent
4) The condition $m_Q^2 \geq 0$ implies the constraint
\be
0\leq k\leq k_{max}~,\label{kmax} \ee on the
integration over the loop momentum $k$, where $k_{max}$ actually depends on 
the kinematics of the process\footnote{For the processes induced by the 
$b \to u$ current, for instance, $k_{max} \simeq m_B/2$ 
(see \cite{Brho,Bpi,Bpipi}). Here $k_{max} \simeq m_D/2$.}
\be
\int\frac{d^3k}{(2\pi)^3}\;\; .
\label{loop}
\ee
\noindent
5) For the weak hadronic current one puts the factor
\be
N_{q} N_{q^\prime}\gamma^\mu (1-\gamma_5)  \;. \label{J} \ee
 The
normalization factor $N_q$ is as follows:
\be
N_q=
\left\{\begin{array}{ccl}
\dd\sqrt{\frac{m_q}{E_q}} &  ~~ &  ({\rm if}~q={\rm constituent~ quark}) \\
&  & \\
1 & ~~ & {\rm (otherwise) \; .}
\end{array}\right .
\label{Nq} \ee 
\noindent 
6) Finally  the amplitude must contain a
colour factor of 3 and a trace over Dirac matrices.

This set of rules can now be applied to the evaluation of the
hadronic matrix element for the decays (\ref{semilep}), 
corresponding to the diagram in fig. \ref{diagram}; the result
is:
\begin{figure}[ht!]
\begin{center}
\epsfig{file=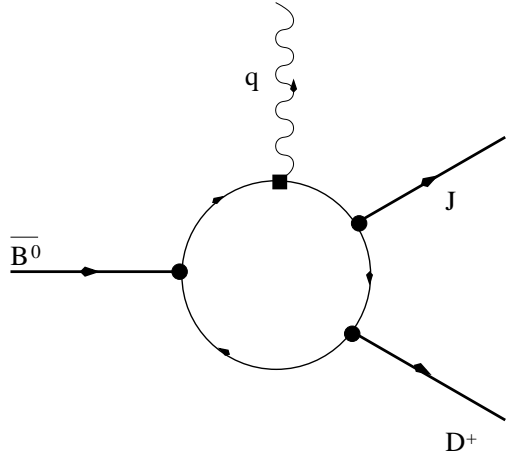,height=6cm}
\end{center}
\caption{Feynman diagram for  $\overline{B^0} \to J~D^+$
semileptonic decay.
$J\equiv J/\psi~,\eta_c$.}
\label{diagram}
\end{figure}
\bea
 &&
\mathcal{A}^\mu=<D^+(p_D)~J|
\bar c \gamma^\mu(1-\gamma_5)b|\overline{B^0}(p_B)>
=
-\sqrt{3} \int\frac{d^3k}{(2\pi)^3} ~\mathcal{N}
~\psi_B(k)~\psi^{\star}_{J}(k)~\psi^{\star}_D(k) \nn\\
 &&
\theta[k_{max}-k]~ Tr\left[ 
\frac{\spur{q_b}+m_b}{2~m_b} \frac{\spur{q_d}+ m_d}{2~m_d} 
\frac{\spur{q_b}-\spur{q}-\spur{p_J}+m_{c^{\prime}}}
{2~m_{c^{\prime}}}
\Gamma \frac{-\spur{q_b}+\spur{q}+ m_c}{2~m_c}
\gamma^{\mu}(1-\gamma_5)
 \right] 
 \label{ampiezza}
\eea
where $\Gamma=\spur{\epsilon^{\star}}~(-i \gamma_5)$ and 
$J \equiv J/\psi(p_J,\epsilon^{\star})~(~\eta_c(p_\eta)~)$ for the outgoing 
$J/\psi~(\eta_c)$ meson. 
Note the appearance of $m_{c^{\prime}}\ne m_c$ resulting 
from (\ref{Alt-Cab2}). 
The factor $\mathcal{N}$ in (\ref{ampiezza}) 
accounts for the normalization of the hadronic current and of the heavy 
mesons:
\bea
&&
\mathcal{N}=\sqrt{\frac{m_c m_b}{E_b (E_b-q^0)}} 
\sqrt{\frac{m_d m_b}{E_d m_B+m_d (m_b-m_d)}} 
\sqrt{\frac{m_d m_{c^{\prime}}}{m_d (m_d-m_{c^{\prime}}) + E_d (E_J-m_B+q^0)}}
\nn\\
&&
\sqrt{\frac{m_c m_{c^{\prime}}}{m_B^2+q^2 - 2 m_c m_{c^{\prime}} + 2 m_d^2
 + m_D^2 - m_J^2 - 2 \left( m_B q^0 - E_d ( E_J - 2 m_B + 2 q^0 ) \right) }} 
\; .
\label{N-norm}
\eea
We introduce the various form factors for the $B \to J/\psi~D$ 
semileptonic decay:
\bea
&&
<D^+(p_D)~J/\psi(p_J,\epsilon^{\star})|
\bar c \gamma^\mu(1-\gamma_5)b|\overline{B^0}(p_B)>=
i~\epsilon^{\star}_\nu \nn\\
&& 
\left\{\left[ m_B^2 C_1 g^{\mu \nu} + p_B^\mu (C_2 p_B^\nu + C_3 q^\nu) + 
 q^\mu (C_4 p_B^\nu + C_5 q^\nu) + p_J^\mu (C_6 p_B^\nu + C_7 q^\nu) 
\right] + \right. \nn\\
&&
+ \epsilon^{\nu\mu\alpha\beta} 
(D_1 p_{J\alpha} q_\beta + D_2 p_{J\alpha} p_{B\beta} + 
D_3 q_\alpha p_{B\beta})+\frac{1}{m_B^2} 
\epsilon^{\mu\sigma\alpha\beta} p_{B\sigma} p_{J\alpha} q_\beta 
(D_4 p_B^\nu + D_5 q^\nu) + \nn\\
&&
\left. \frac{1}{m_B^2} \epsilon^{\nu\sigma\alpha\beta} 
p_{B\sigma} p_{J\alpha} q_\beta 
(D_6 p_B^\mu + D_7 p_J^\mu + D_8 q^\mu)\ \right\}
\;\; , \label{BJD} \eea
\noindent
where $q=p_B-p_D-p_J$.
Following \cite{lee} we introduce also the form factors for the 
$B \to \eta_c~D$ semileptonic decay as
follows:  
 
\bea
&&
<D^+(p_D)~\eta_c(p_\eta)|
\bar c \gamma^\mu(1-\gamma_5)b|\overline{B^0}(p_B)>=\nn\\
&&=
i~w_+~(p_D+p_\eta)^{\mu}~+~i~w_-~(p_\eta-p_D)^{\mu}~+~{i}~r~q^\mu~+~
2~h~\epsilon^{\mu\alpha\beta\delta}~p_\alpha p_{\eta
\beta}p_{D\delta} \;\; . \label{BetacD} \eea

\noindent It is useful to introduce the following variables:
\begin{eqnarray*}
s~&=&~(p_D+p_J)^2 \\
t~&=&~(p-p_J)^2 \\ 
u~&=&~(p-p_D)^2~,
\end{eqnarray*}
that satisfy
\begin{equation}
s+t+u=q^2+m_B^2+m_D^2+m_J^2~.
\end{equation}

The form factors
$w_\pm,~r,~h,~C_i,~D_j$ (i=1,...,7;~j=1,...,8) are functions of three 
independent variables. One
can  choose as independent variables $s,~q^2,~t$ or,
alternatively, $s,~q^2,~E_J$, where $E_J$ 
is the $J/\psi~(\eta_c)$
energy in the $B$-meson rest frame. The relations between the two sets
of invariants are:
\begin{eqnarray}\label{rel}
t&=&~m^2_B~+~m^2_J~~-~2m_B~E_J \nn\\
q^2&=&~s~+~m_B^2~-~2~m_B~(E_D~+~E_J)~.
\end{eqnarray}
The kinematical range is as follows:
\begin{eqnarray}\label{rel1}
m_D^2 \leq & t & \leq (m_B-m_J)^2 \nn\\
0 \leq & q^2 & \leq (m_B-m_J-\sqrt{t})^2 \nn\\
s_{min} \leq & s & \leq s_{max} \; ,
\end{eqnarray}
where
\begin{eqnarray}
&&
s_{min/max} = 
\frac{1}{4 t}
\left[ \left( m_B^2 + m_D^2 - m_J^2 - q^2 \right)^2 - 
\left( \sqrt{\lambda(t,m_J^2,m_B^2)} \pm \sqrt{\lambda(m_D^2,t,q^2)}
\right)^2 
\right] \; ; \nn\\
&&
\lambda(x,y,z)=~x^2+y^2+z^2-2 (x y+y z+x z) \; .
\label{p}
\end{eqnarray}
From eq. (\ref{ampiezza})  one can extract the different
form factors by multiplying $\mathcal{A}^\mu$ by appropriate momenta 
(see the Appendix for explicit expressions of  all the form factors).

The calculation of the trace in
(\ref{ampiezza}) is straightforward and is similar to those
performed in  \cite{Brho, Bpi, Bpipi} for similar processes.
The evaluation of the integral is 
even simpler,
because, although  the three-body decay kinematics is rather involved, 
all the quarks in the loop of fig. \ref{diagram} are constituent and there 
is no 
light quark propagating.
The integration can be performed numerically but, unlike the semileptonic
decays with two pions in the final state \cite{Bpipi}, 
here the loop integration domain is not genuinely
three-dimensional, due to the smallness of phase space. In fact, 
the calculation becomes simpler by inserting the wave functions 
$\psi_H(k)~(H=~D,~J/\psi,~\eta_c)$ as 
they are in the $H$-meson rest frame in the relevant formulae, 
which is an approximation we perform and is justified by the small value of
the outgoing mesons' $4$-momenta in the processes 
(\ref{semilep}). This approximation has been already 
incorporated into the formulae of the Appendix
\footnote{The form factors in (\ref{BJD},\ref{BetacD})
do not differ significantly from their actual value in the kinematical range 
(\ref{rel1}) and within this approximation.}.

An important point to be stressed is the kinematical range in
which the predictions of the present model are reliable. We cannot
pretend to extend our analysis to very small meson momenta for the
following reasons:  first, as discussed in \cite{Bpi,Bpipi}, when
the outgoing meson momenta are small, 
the results of the model become strongly
dependent on a numerical input of our calculation, {\it i.e.} the value
of the light quark mass $m_d$. The numerical value of $m_d$ cannot
be fixed adequately because the values of the quark masses were
fitted from the heavy meson spectrum, which is not very sensitive
to the light quark mass (for more details see \cite{pietroni}). Therefore 
the value of $m_d$ has a theoretical uncertainty and the parameter $m_d$ is the main 
source of error for the present calculation. We set the value of this 
parameter to the scale $\Lambda_{\rm QCD}$\footnote{In \cite{Bpipi} the 
low-lying $\rho$-resonance provides the model with a cut-off at small 
$s$ in the $B\to\pi\pi$ semileptonic decay form factors; here such a {\it natural} 
hadronic scale is absent.}. 

\begin{figure}[ht!]
\begin{center}
\epsfig{file=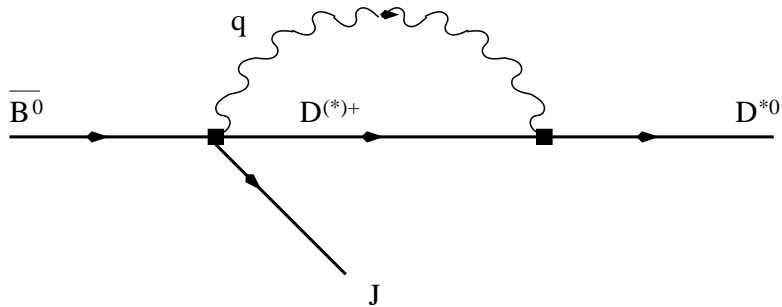,height=4cm}
\end{center}
\caption{Feynman diagram for 
$\overline{B^0} \to J~D^{\star 0}$
nonleptonic decay in the QCD 
relativistic potential model. $J\equiv J/\psi~,\eta_c$.} 
\label{diagram2}
\end{figure}

Moreover the role of pole
diagrams such as those studied in \cite{ELY} becomes relevant.
These diagrams cannot be accounted for by the  present scheme,
which at most can be used to model a continuum of states,
according to the quark-hadron duality ideas. The low-lying
resonances, such as those studied in \cite{ELY} should be added
separately\footnote{This is the reason why in \cite{Bpi} the $B^*$
pole of the $B\to\pi$ form factor is not reproduced in the $|\vec
p_\pi|\to 0$ region. The same remark holds for \cite{Bpipi} about the
resonances encountered at small $s$ in the $B\to\pi\pi$ semileptonic decay 
form factors, such as the $\rho$-resonance.}. We expect a large
contribution from the $D^{\star}$ (see the discussion in
\cite{ELY}). It is worthwhile to stress that in the present model the resonant 
production of a pion occurs through the long distance contribution 
depicted in fig. \ref{diagram2} with $D^{(\star) +}$ mesons 
as intermediate states; 
all the diagrams are calculable in the Heavy Meson Chiral Lagrangean 
\cite{casalbuoni} and they are found to be of the same order of those calculated 
in \cite{ELY}.   
\par
In principle the partial widths 
$\Gamma(\overline{B^0} \to J~D^+~\ell \bar\nu_\ell)$ can 
be used to extract the relevant $|V_{cb}|$ CKM matrix element: due to the smallness 
of the phase space, the theoretical predictions are not expected to strongly depend 
on the specific model employed to achieve the final results. 
We find $\mathcal{BR}(\overline{B^0} \to 
J/\psi~(\eta_c)~D^+~\ell~\nu_\ell) \simeq 10^{-13}$, which 
is of course unmeasurable in the foreseeable future.

\begin{figure}[ht!]
\begin{center}
\epsfig{file=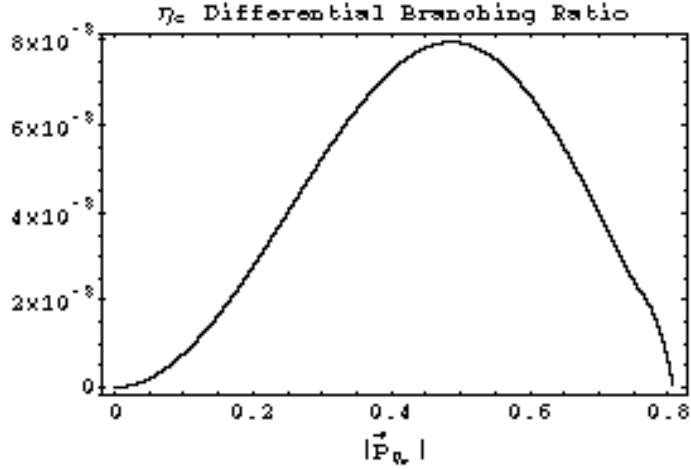,height=6.2cm}
\end{center}
\caption{Differential branching ratio, in ${\rm GeV}^{-1}$, 
for $\overline{B^0} \to \eta_c~D^+~\pi^-$. The momentum is in GeV.}
\label{BF-eta}
\end{figure}

\begin{figure}[ht!]
\begin{center}
\epsfig{file=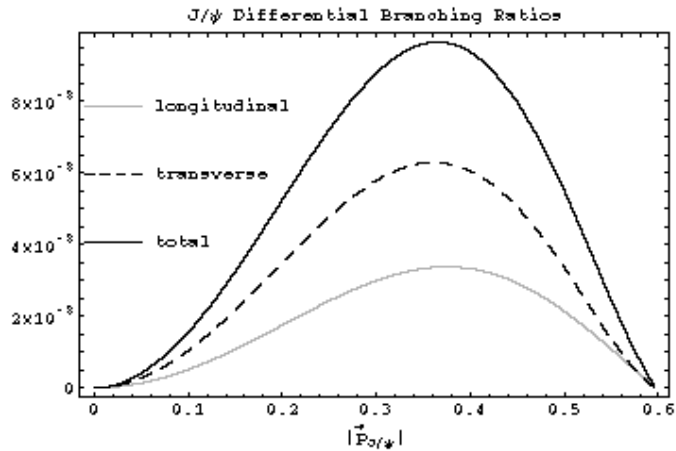,height=6.2cm} 
\end{center}
\caption{Differential branching ratio, in ${\rm GeV}^{-1}$, 
for $\overline{B^0} \to J/\Psi~D^+~\pi^-$. The momentum is in GeV.}
\label{BF-JPsi}
\end{figure}

Instead, we calculate the relevant formulae of the 
$\overline{B^0} \to J~D^+~\pi^-$ nonleptonic decays. This 
channel is a background to  
$\overline{B^0} \to J~D^{\star 0}$ followed by 
the (almost on-shell) decay $D^{\star 0} \to D^+~\pi^-$.
The relevant amplitudes follow from eqs. (\ref{BJD}) and (\ref{BetacD}). 
Numerically we get (for $m_d=300~{\rm MeV} \simeq \Lambda_{\rm QCD}$):
\bea
&&
{\cal BR}( \overline{B^0} \to J/\psi~D^+~\pi^- ) =
\left\{
\begin{array}{ccl}
1.98 \times 10^{-8} &  ~~ &  {\rm transverse~polarization} \\
 & & \\
1.09 \times 10^{-8} &  ~~ &  {\rm longitudinal~polarization \; ,}
\end{array} 
\right. \nn\\
&& \nn\\
&&
{\cal BR}( \overline{B^0} \to \eta_c~D^+~\pi^- ) = 3.54 \times 10^{-8} \; .
\label{Br-Ra} 
\eea 
It is also interesting to compute the  differential branching ratios: 
\be
\frac{d \mathcal{BR}}{d|\vec p_J|} = 
\frac{\tau_{B^0} f_\pi^2 |V_{cb} V_{ud}^{\star}|^2 G_F^2}{256 \pi^3 m_B^2} 
\frac{|\vec p_J|}{E_J} 
\int_{s_{min}}^{s_{max}} ds~|q\mathcal{A}(t,q^2,s)|^2 \; ,
\label{BF}
\ee
where $q^2=m_\pi^2,~f_\pi=132$ MeV $\tau_{B^0}=1.6$ 
ps$,~V_{cb}=0.040,~V_{ud}=1-\lambda^2/2,~\lambda=0.22,~m_B=5.279$ GeV$,~G_F$ 
is the Fermi constant and $s_{min/max}$ are as in 
(\ref{p}). $\mathcal{A}^\mu$ is the relevant amplitude for the nonleptonic 
decay $\overline{B^0}\to J/\psi~(\eta_c)~D^+~\pi^-$ given in eq. 
(\ref{ampiezza}).
The differential branching ratios
for the nonleptonic decays we have 
studied are plotted in figs. \ref{BF-eta} and \ref{BF-JPsi}.


We can therefore conclude that from an experimental point of view
the $B-$meson nonleptonic decay channel with three charm quarks in the
final state represents a (barely) 
interesting process. We have investigated the 
$\overline{B^0} \to J/\psi~(\eta_c)~D^+~\ell~\nu_\ell$ semileptonic decays
and the $\overline{B^0}\to J/\psi~(\eta_c)~D^+~\pi^-$ nonresonant 
nonleptonic decay channels by using the factorization approximation 
and the $\overline{B^0}\to J/\psi~(\eta_c)~D^+$ semileptonic decay 
form factors. The calculation has been performed in the framework of a 
QCD relativistic potential model. The branching ratios for these decays are: 
$\mathcal{BR}(\overline{B^0}\to J/\psi~(\eta_c)~D^+~\ell~\nu_\ell \simeq 10^{-13}$, 
$\mathcal{BR}(\overline{B^0}\to J/\psi~D^+~\pi^-)=3.07 \times 10^{-8}$ and 
$\mathcal{BR}(\overline{B^0}\to \eta_c~D^+~\pi^-)=3.54 \times 10^{-8}$. 
These theoretical findings provide us with 
an order of magnitude for those processes and, 
in that respect, they do not seem to account for the slow $J/\psi$ hump as 
indicated by the preliminary results of CLEO and Belle collaborations. 
Since the charmonium spectrum will be extensively studied at the 
$B$-factories in the near future, it is important to confirm whether the 
slow $J/\psi$ hump exists. In that respect, a refined measurement is 
needed. If the hump persists, it will be hard to find a consistent explanation 
within the conventional models. Thus a new scenario, like those discussed 
in \cite{brod,hou,ELY}, could be applicable.
\vskip 0.2cm
\par\noindent{\bf Acknowledgments} 
This work has been supported in part by the Israel-USA
Binational Foundation and by the Israel Science Foundation.
The research of G.E. has been supported in part by the Fund
for the Promotion of Research at the Technion.
\newpage
\begin{large}
{\bf Appendix}
\end{large}
\par\noindent 
From eq. (\ref{ampiezza})  one can extract the different
form factors by multiplying $\mathcal{A}^\mu$ by appropriate momenta. One
gets\footnote{$D_j=0~(j=4,...,8)$ is a consequence of the
$H\bar q Q~(H=B,~D,~J/\psi,~\eta_c)$ couplings introduced in our model 
and of the nature of the quarks involved (all constituents): 
there is no way to generate higher powers in the meson $4$-momenta 
in the loop of fig. \ref{diagram}.}:
\bea 
&&
C_1=\int_0^{k_{max}}\frac{dk~k^2}{2 \pi^2}~\mathcal{N}
~\frac{\psi_B(k)~\psi^{\star}_{J}(k)~\psi^{\star}_D(k)}
{8 \sqrt{6} m_B^2 m_d^2 m_c m_{c^{\prime}}^2 m_b} 
\left[ m_d (m_d -m_{c^{\prime}}) + E_d (E_J-m_B+q^0)\right] \nn\\
&&
\left\{ 2 E_d^2 m_B (m_B+q^0)+
E_d \left[
-m_B^3-2 m_B^2 q^0+(m_b-m_d)(m_c-m_{c^{\prime}})q^0 +\right. \right. \nn\\
&&
E_J \left( -(m_b-m_d)(m_c+m_d)+m_B (m_B+q^0) \right)+ \nn\\
&&
\left. 
m_B \left( m_b (m_c-m_{c^{\prime}}) +m _c m_{c^{\prime}} +
     2 m_d (m_b-m_c+m_{c^{\prime}}) -3 m_d^2 -q^2 \right) \right] + \nn\\
&&
m_d \left[ m_B^2 (m_c-m_{c^{\prime}}+m_d) + E_J m_B (m_b-m_c-2 m_d) + 
\right. \nn\\
&&
m_d (m_d+m_c) (m_d-m_{c^{\prime}}) +m_B q^0 (m_c - m_{c^{\prime}}+2 m_d) 
+\nn\\
&&
\left. \left.
E_J (m_b-m_d) q^0 + m_d q^2 - 
 m_b \left( (m_d-m_{c^{\prime}}) (m_d+m_c)+m_B (m_B+2 q^0) +q^2 
\right) \right] \right\} 
\; , \nn\\
&& \nn\\ &&
C_2=-2\int_0^{k_{max}}\frac{dk~k^2}{2 \pi^2}~\mathcal{N}
~\frac{\psi_B(k)~\psi^{\star}_{J}(k)~\psi^{\star}_D(k)}
{8 \sqrt{6} m_B^2 m_d^2 m_c m_{c^{\prime}}^2 m_b} 
\left[ m_d (m_d -m_{c^{\prime}}) + E_d (E_J-m_B+q^0)\right]\nn\\
&& 
\left( E_d-m_B \right)
\left[ m_B m_d (m_b-m_c-2 m_d) + E_d (m_B^2-(m_b-m_d) (m_c+m_d)) \right]
\; , \nn\\
&& \nn\\ &&
C_3=m_B\int_0^{k_{max}}\frac{dk~k^2}{2 \pi^2}~\mathcal{N}
~\frac{\psi_B(k)~\psi^{\star}_{J}(k)~\psi^{\star}_D(k)}
{8 \sqrt{3} m_B^2 m_d^2 m_c m_{c^{\prime}}^2 m_b} 
\left[ m_d (m_d -m_{c^{\prime}}) + E_d (E_J-m_B+q^0)\right]\nn\\
&& 
\left[ m_B m_d (2 m_b-m_c-m_{c^{\prime}}-4 m_d) + 
E_d (2 m_B^2-(m_b-m_d) (m_c+m_{c^{\prime}}+2 m_d)) \right] 
\; , \nn\\ 
&& \nn\\ &&
C_4=-m_B\int_0^{k_{max}}\frac{dk~k^2}{2 \pi^2}~\mathcal{N}
~\frac{\psi_B(k)~\psi^{\star}_{J}(k)~\psi^{\star}_D(k)}
{8 \sqrt{3} m_B^2 m_d^2 m_c m_{c^{\prime}}^2 m_b} 
\left[ m_d (m_d -m_{c^{\prime}}) + E_d (E_J-m_B+q^0)\right]\nn\\
&& 
\left[ -2 E_d E_b m_B+ m_B m_d (-2 m_b+m_c-m_{c^{\prime}}+2 m_d) + 
E_d ((m_b-m_d) (m_c-m_{c^{\prime}}+2 m_d)) \right]
\; , \nn\\
&& \nn\\ &&
C_5=2 m_B^2 \int_0^{k_{max}}\frac{dk~k^2}{2 \pi^2}~\mathcal{N}
~\frac{\psi_B(k)~\psi^{\star}_{J}(k)~\psi^{\star}_D(k)}
{8 \sqrt{3} m_B^2 m_d^2 m_c m_{c^{\prime}}^2 m_b} 
\left[ m_d (m_d -m_{c^{\prime}}) + E_d (E_J-m_B+q^0)\right]\nn\\
&& 
\left[ E_d m_B + m_d (m_b-m_d)\right] \; , \nn\\
&& \nn\\ &&
C_7=C_6=-\frac{1}{2} C_5 \; , \nn\\
&& \nn\\ &&
D_1=m_B^2 \int_0^{k_{max}}\frac{dk~k^2}{2 \pi^2}~\mathcal{N}
~\frac{\psi_B(k)~\psi^{\star}_{J}(k)~\psi^{\star}_D(k)}
{8 \sqrt{3} m_B^2 m_d^2 m_c m_{c^{\prime}}^2 m_b} 
\left[ m_d (m_d -m_{c^{\prime}}) + E_d (E_J-m_B+q^0)\right]\nn\\
&& 
\left[ E_d m_B + m_d (m_b-m_d)\right] \; , \nn\\
&& \nn\\ &&
D_2=m_B \int_0^{k_{max}}\frac{dk~k^2}{2 \pi^2}~\mathcal{N}
~\frac{\psi_B(k)~\psi^{\star}_{J}(k)~\psi^{\star}_D(k)}
{8 \sqrt{3} m_B^2 m_d^2 m_c m_{c^{\prime}}^2 m_b} 
\left[ m_d (m_d -m_{c^{\prime}}) + E_d (E_J-m_B+q^0)\right]\nn\\
&& 
\left[ m_B m_d (m_b-m_c-2 m_d) + 
E_d \left( m_B^2 - (m_b-m_d) (m_c+m_d) \right) \right] \; , \nn\\
&& \nn\\ &&
D_3=-m_B \int_0^{k_{max}}\frac{dk~k^2}{2 \pi^2}~\mathcal{N}
~\frac{\psi_B(k)~\psi^{\star}_{J}(k)~\psi^{\star}_D(k)}
{8 \sqrt{3} m_B^2 m_d^2 m_c m_{c^{\prime}}^2 m_b} 
\left[ m_d (m_d -m_{c^{\prime}}) + E_d (E_J-m_B+q^0)\right]\nn\\
&& 
\left[ (m_c-m_{c^{\prime}}) 
\left( E_d (m_b-m_d) + m_B m_d \right) \right] \; , \nn\\
&& \nn\\ &&
D_4=D_5=D_6=D_7=D_8=0 \; , \nn\\
&& \nn\\ &&
w_+=-\frac{m_B}{2} \int_0^{k_{max}}\frac{dk~k^2}{2 \pi^2}~\mathcal{N}
~\frac{\psi_B(k)~\psi^{\star}_{J}(k)~\psi^{\star}_D(k)}
{8 \sqrt{3} m_B^2 m_d^2 m_c m_{c^{\prime}}^2 m_b} 
\left[ m_d (m_d -m_{c^{\prime}}) + E_d (E_J-m_B+q^0)\right]\nn\\
&&
\left\{ -4 E_d^2 (m_b-m_d) (m_B-q^0) + m_B m_d \right. \nn\\
&&
\left[ m_B^2 -2 m_{c^{\prime}} (m_c+2 m_d)+ m_d (3 m_c+m_d)
+ m_b (-m_c+2 m_{c^{\prime}} + m_d) + 2 (E_J q^0 + q^2) \right] + \nn\\
&&
E_d \left[ -m_B^2 (m_c-2 m_{c^{\prime}}+4 m_d) -2 m_d 
(-m_c m_{c^{\prime}} + m_c m_d +m_d^2+E_J q^0+q^2) + \right. \nn\\
&&
\left. \left.
7 m_B m_d q^0+m_b \left( m_B^2-3 m_B q^0+2 
(-m_c m_{c^{\prime}} + m_c m_d +m_d^2+E_J q^0+q^2) \right) \right] \right\}
 \; , \nn\\
&& \nn\\ &&
w_-=-\frac{m_B^2}{2} \int_0^{k_{max}}\frac{dk~k^2}{2 \pi^2}~\mathcal{N}
~\frac{\psi_B(k)~\psi^{\star}_{J}(k)~\psi^{\star}_D(k)}
{8 \sqrt{3} m_B^2 m_d^2 m_c m_{c^{\prime}}^2 m_b} 
\left[ m_d (m_d -m_{c^{\prime}}) + E_d (E_J-m_B+q^0)\right]\nn\\
&&
\left[ (m_b-m_d) \left( E_d (q^0-m_B)+m_d(m_d+m_c) \right)
+ m_B \left( m_d(q^0-m_B)+E_d(m_c+m_d) \right) \right]  
\; , \nn\\
&& \nn\\ &&
r  =m_B \int_0^{k_{max}}\frac{dk~k^2}{2 \pi^2}~\mathcal{N}
~\frac{\psi_B(k)~\psi^{\star}_{J}(k)~\psi^{\star}_D(k)}
{8 \sqrt{3} m_B^2 m_d^2 m_c m_{c^{\prime}}^2 m_b} 
\left[ m_d (m_d -m_{c^{\prime}}) + E_d (E_J-m_B+q^0)\right]\nn\\
&&
(E_d m_b+E_b m_d) \left[ m_B q^0+(m_B -q^0) (E_J+2 E_d-m_B) -q^2 +
(m_c+m_d) (m_{c^{\prime}} -m_d) \right] 
 \; , \nn\\
&& \nn\\ &&
h  = \frac{m_B}{2} \int_0^{k_{max}}\frac{dk~k^2}{2 \pi^2}~\mathcal{N}
~\frac{\psi_B(k)~\psi^{\star}_{J}(k)~\psi^{\star}_D(k)}
{8 \sqrt{3} m_B^2 m_d^2 m_c m_{c^{\prime}}^2 m_b} 
\left[ m_d (m_d -m_{c^{\prime}}) + E_d (E_J-m_B+q^0)\right]\nn\\
&&
\left[ E_d (m_b-m_d) + m_B m_d \right]\nn \; .
\label{ff4} 
\eea


\end{document}